\documentstyle[prl,eqsecnum,aps,twocolumn]{revtex}
\catcode`\@=11

\def\maketitle2{\par % Uses \twocolumn[\@maketitle2].
\begingroup
\let\cite\@bylinecite
\def\thefootnote{\fnsymbol{footnote}}%
\twocolumn[\@maketitle2\vskip2pc]%
\thispagestyle{plain}\@thanks
\endgroup
\def\thefootnote{\arabic{footnote}}%
\setcounter{footnote}{0}%
\let\maketitle2\relax \let\@maketitle2\relax
\let\@thanks\relax \let\@authoraddress\relax \let\@title\relax
\let\@date\relax \let\thanks\relax \let\@abstract\relax 
\let\@pacs\relax}

\def\abstract#1{\gdef\@abstract{{\par % Store abstract text. 
\bgroup
\ifdim\prevdepth=-1000pt \prevdepth0pt\fi
\hsize\columnwidth
\dimen0=-\prevdepth \advance\dimen0 by17.5pt \nointerlineskip
\small\vrule width 0pt height\dimen0 \relax}{~~}#1\egroup}}

\def\pacs#1{\gdef\@pacs{{\par % Store PACS numbers as \@pacs.
\bgroup
\hsize\columnwidth \parindent0pt
\ifdim\prevdepth=-1000pt \prevdepth0pt\fi
\dimen0=-\prevdepth \advance\dimen0 by20pt\nointerlineskip
\egroup} PACS numbers:~#1}}

\def\@maketitle2{% Puts \@abstract and \@pacs in a {list}.
\@preprint
\@title
\ifdim\prevdepth=-1000pt \prevdepth0pt\fi
\@authoraddress
\@date
\begin{list}{}{\leftmargin=0.10753\textwidth \rightmargin=\leftmargin
\itemsep=1pc\partopsep=-1pc}
\item\@abstract
\item\@pacs
\end{list}
}

\catcode`\@=12

\begin{document}
\title{A Prosaic Explanation for the Anomalous Accelerations Seen in
Distant Spacecraft}
\author{Edward M. Murphy\cite{byline}}
\address{The Johns Hopkins University\\Dept. of Physics and
Astronomy\\
3400 N. Charles Street\\Baltimore, MD 21218}
\date{September 21, 1998}

\abstract{
Anderson, et al. \cite{anderson98} have recently reported
the discovery of an apparent anomalous, weak, long-range acceleration
in the Pioneer 10/11 and Ulysses spacecraft.  I believe that this
result can be explained by non-isotropic radiative cooling of the
spacecraft electronics through passive radiators on the spacecraft
surface.  These radiators are preferentially placed on the anti-solar
side of the spacecraft to avoid heating by solar radiation.  The power
transmitted through these radiator panels can explain the observed
acceleration within the observational errors.}

 
%***************************************************
\maketitle2              %****MACROS***************
\narrowtext              %*************************
%*******************************************
 
\section{Introduction}

Anderson, et al. \cite{anderson98} have recently modeled
the accelerations acting on the Pioneer 10, Pioneer 11, Ulysses, and Galileo
spacecraft.  They find an anomalous, excess acceleration of
$\rm 8.5\times10^{-8}\;cm\;s^{-2}$ directed towards the Sun.  They have ruled
out excess gravitational forces due to the Galaxy and unidentified
planetesimals, errors in the orbital and rotational parameters of the
Earth, spacecraft gas leaks, and errors in the planetary ephemeris as
explanations for the acceleration.  In addition, the authors rule out
radiation pressure from thermal radiation generated by the spacecraft
radioisotope thermoelectric generators (RTGs).  Anderson et al. assume
that the thermal radiation generated by the RTGs is
isotropically radiated and results in no net force on the
spacecraft.  However, I believe that this assumption overlooks the
fact that the electrical energy produced by the RTGs is dissipated in a
non-isotropic manner.

Spacecraft designed to travel beyond the inner solar system cannot
rely on currently available solar cells to provide power, as the size
of the solar arrays would be prohibitively large.  Therefore, missions
to the outer solar system have used RTGs to provide power.  RTGs rely
on the thermal energy generated by the radioactive decay of Pu$^{238}$
to heat a semiconductor junction which generates an electrical current.
RTGs have an electrical conversion efficiency of a few percent
\cite{piscane94}.  For example, the Ulysses RTGs generate 4500
W of thermal power and produce 280 W of electrical power (at the
beginning of the mission).  The available power decreases with time
due to degradation in the semiconductor junction and, to a much lesser
degree, the decay of the Pu$^{238}$ \cite{piscane94}.

The excess thermal power generated by the RTGs is dumped radiatively
by cooling fins located on the outer surface of the cylindrical RTG
structure (this is true of all the spacecraft considered here).  The
geometry of the fins is complex and the thermal radiation dissipated
from the surface will not be isotropic.  However, a cursory
examination of the Pioneer and Ulysses RTG designs shows that they are
cylindrically symmetric.  Even though it is not isotropic, the escaping
radiation will not impart a net force on the spacecraft since 
it is dissipated symmetrically.

The same is not true of the electrical energy created by the RTGs.
The electrical energy is transported to a main power bus from where it
is distributed to the individual subsystems of the satellite to
provide power for operating the electronics.  These electronics are
typically found in a single large electronics bay (which contains most
of the essential systems) and science instruments distributed
throughout and around the spacecraft.  To prevent the electronics from
overheating, the waste heat dissipated by the electronics is radiated
from the spacecraft by surface radiators.  In most cases, radiators
are located on the anti-solar side of the spacecraft to prevent the
panels from being heated by solar radiation.  Because the radiator
panels are preferentially located on the anti-solar side of the
spacecraft, their radiation will cause an acceleration of the
spacecraft toward the Sun.

From conservation of momentum arguments, it is easy to show that the
acceleration, $a_P$, produced by an amount of radiated power, $P$,
is $a_{P}=P\:(m\:c)^{-1}$ where $m$ is the mass of the spacecraft and
$c$ is the speed of light.  This assumes that the radiated power is
tightly collimated (i.e. it carries all the momentum in a single
direction).  In fact, however, the radiation from a flat plate is
spread over 2$\pi$ steradians.  In the case of a flat Lambertian
source (i.e. one in which the intensity is independent of viewing
angle \cite{boyd83}), the momentum carried away perpendicular to the
plate surface will be 2/3 of the total.

\section{Ulysses}

The Ulysses spacecraft is spin stabilized with the rotation axis
pointing approximately toward the Earth (and the Sun when the
spacecraft is near aphelion).  The anti-Earth (anti-solar) side is
always in the spacecraft shadow.  The majority of the electrical
components in the Ulysses spacecraft are located in a single thermal
enclosure \cite{standley98}.  The waste heat from the electronics is
radiated through a large, flat radiator panel on the anti-solar side
of the spacecraft.  The interior electronics radiate their heat to the
panel, which in turn radiates the heat into space.  In addition, the
traveling wave tube amplifier (which dissipates 43 W alone) is directly
thermally coupled to the surface 
radiator.  Except for the anti-solar side, the spacecraft is covered
in multi-layer insulation (MLI) blankets.  A large 1.65 meter diameter
antenna covers most of the Earth facing side.

A power budget for the Ulysses spacecraft for January 1998
\cite{standley98} indicates that Ulysses' systems are drawing 
$231\pm3$ W of electrical power.  Of this, I calculate that $27\pm10$
W is dissipated by scientific instruments and heaters outside the main
thermal enclosure and another 20 W is radiated by the transmitter.
Therefore, in a steady state, the main thermal enclosure must radiate
$184\pm13$ W of power.  Because the error estimates are systematic,
rather than statistical, I have added them directly rather than in
quadrature.  Some fraction of this power will escape through the MLI
thermal blankets and the remainder will be radiated through the large
surface radiator on the anti-solar side of the spacecraft.  MLI
blankets typically radiate 8 W m$^{-2}$ \cite{piscane94} into
space. Only the blankets on the side of the spacecraft will radiate
internal heat because the solar facing blankets of Ulysses
allow a net input of heat into the thermal enclosure due to solar
heating, though this input power is small ($\sim$ 2 W) compared to
electrical power 
when the spacecraft is at aphelion.  When near perihelion, Ulysses
compensates for the excess input solar heating by dumping excess
electrical energy into resistors on the outside of the spacecraft.
About 50 W of excess electrical power is dumped when the spacecraft is
near perihelion \cite{standley98}. 

I calculate that there are $3.0\pm1.0\:{\rm m}^{2}$ of MLI blankets on the
sides of the main thermal enclosure of Ulysses resulting in $24\pm8$ W
escaping through the MLI blankets.  This implies that the total power
radiated through the spacecraft radiator on the anti-solar side of
Ulysses is ($160\pm21$) W.  The acceleration produced by this power is
$a_{P}=(10.3\pm1.3)\times 10^{-8}\:{\rm cm}\:{\rm s}^{-2}$
assuming a spacecraft mass of 345 kg and that the radiator is a
Lambertian source (2/3 of the momentum is carried away perpendicular
to the radiator).  Since the radiator faces away from the Sun, the
direction of this acceleration is toward the Sun.  This
matches, to within the errors, the anomalous acceleration reported by
Anderson et al. \cite{anderson98} for Ulysses of
$a_{P}=(12\pm3)\times10^{-8}\: {\rm cm}\:{\rm s}^{-2}$.

\section{Pioneer 10 and 11}

Toward the ends of their missions, the Pioneer 10/11 spacecraft were
drawing 80 W of electrical power \cite{anderson98} from their RTGs,
which was sufficient to power the essential spacecraft systems and
possibly one or two scientific instruments.  Of this, 9 W is
transmitted as RF power \cite{anderson98}.  The essential electrical
systems are located in a cylindrical hub beneath the high-gain
antenna.  The waste heat generated by the electronics is radiated from
a series of fins on the anti-solar side of the spacecraft.  Since the
majority of the science instruments have been turned off, essentially
all of the 71 W of internally dissipated electrical power is radiated
from the fins.  Solar heating is negligible at the Pioneers' distance
from the Sun.  Assuming that the current mass of Pioneer 10/11 is 250
kg, the radiated power generates an $a_{P}=6.3\times10^{-8}\:
{\rm cm}\:{\rm s}^{-2}$ again assuming a Lambertian source.  However,
the radiator 
fins of the back of the Pioneer spacecraft are highly non-Lambertian
sources.  In fact, the fins are likely to collimate the outgoing
radiation to a significant degree.  If the radiation were fully
collimated, the resulting acceleration would be
$a_{P}=9.5\times10^{-8}\: {\rm cm}\:{\rm s}^{-2}$.  The actual value
is likely to lie 
between these extremes. These estimates of the acceleration due to
radiative cooling closely match the $a_{P}=8.5\times10^{-8}\;{\rm
cm}\;{\rm s}^{-2}$ reported by Anderson, et al. \cite
{anderson98} for the Pioneer 10 and 11 spacecraft.

The essential electrical systems must remain powered at all times.
Although the thermal power output of the RTGs is expected to decrease
with time, the power drawn by the essential spacecraft electronics is
nearly constant and, therefore, the acceleration imparted by the
thermal radiation from the spacecraft radiators should also be
constant with time.  As the missions have progressed, various science
instruments have been turned on and off.  Most of these instruments
are located on the periphery of the spacecraft, or on long booms
reaching 3 meters or more from the spacecraft.  These instruments are,
typically, small enough that they radiate their power isotropically
and do not employ a radiator system.  Therefore, as the the science
instruments are cycled on and off, there should be little effect on
the net acceleration.

\section{Conclusions}

I have shown that the most likely explanation for the anomalous
accelerations found by Anderson et al. in the Pioneer 10/11 and
Ulysses spacecraft is radiation pressure from spacecraft radiators
which prevent the buildup of heat in the electronics systems.
A more detailed analysis of the thermal designs of the spacecraft,
including an examination of their power requirements with time,
would allow us to reduce the associated errors and would make it
possible to search for additional accelerations in the Anderson et
al. data at a level of $a_{P}=1-2\times10^{-8}\;{\rm cm}\;{\rm s}^{-2}$.

\acknowledgements

I wish to thank Shaun Standley of the Ulysses Project at JPL for very
informative conversations concerning the Ulysses thermal design and
power budget.  I
also wish to thank Scott Friedman and Alexandra Cha for helpful
conversations.
This work was supported by NASA contract NAS5-32985 (FUSE).

\end{document}